\documentclass[apj]{emulateapj} 

\usepackage{times,amsmath,amssymb,amstext,graphicx,epstopdf,hyperref,subfigure,epsfig,aas_macros,natbib}

\usepackage{color}
\definecolor{myColor}{rgb}{0.9,0.9,0.9}  

\submitted{Accepted for publication in The Astrophysical Journal Letters}

\begin{document}
\renewcommand\bottomfraction{.9}
\title{Towards Chemical Constraints on Hot Jupiter Migration}
\shorttitle{Chemical Constraints on Hot Jupiter Migration} 
\author{Nikku Madhusudhan\altaffilmark{1}, Mustafa A. Amin\altaffilmark{1,2}, Grant M. Kennedy\altaffilmark{1}}
\altaffiltext{1}{Institute of Astronomy, University of Cambridge, Cambridge CB3 0HA, United Kingdom {\tt nmadhu@ast.cam.ac.uk}}
\altaffiltext{2}{Kavli Institute for Cosmology, University of Cambridge, Madingley Road, Cambridge CB3 0HA, UK}

\begin{abstract}
  The origin of hot Jupiters --- gas giant exoplanets orbiting very close to their host
  stars --- is a long-standing puzzle. Planet formation theories suggest that such
  planets are unlikely to have formed in-situ but instead may have formed at large
  orbital separations beyond the snow line and migrated inward to their present
  orbits. Two competing hypotheses suggest that the planets migrated either through
  interaction with the protoplanetary disk during their formation, or by disk-free
  mechanisms such as gravitational interactions with a third body. Observations of
  eccentricities and spin-orbit misalignments of hot Jupiter systems have
  been unable to differentiate between the two hypotheses. In the present work, we
  suggest that chemical depletions in hot Jupiter atmospheres might be able to constrain
  their migration mechanisms. We find that sub-solar carbon and oxygen abundances in
  Jovian-mass hot Jupiters around Sun-like stars are hard to explain by disk
  migration. Instead, such abundances are more readily explained by giant planets 
  forming at large orbital separations, either by core accretion or gravitational instability, and 
  migrating to close-in orbits via disk-free mechanisms involving dynamical 
  encounters. Such planets also contain solar or super-solar C/O ratios. On the contrary, hot
  Jupiters with super-solar O and C abundances can be explained by a variety of
  formation-migration pathways which, however, lead to solar or sub-solar C/O 
  ratios. Current estimates of low oxygen abundances in hot Jupiter atmospheres may be 
  indicative of disk-free migration mechanisms. We discuss open questions 
  in this area which future studies will need to investigate. 
\end{abstract} 


\section{Introduction}
\label{sec:intro}

Gas giant planets are thought to form in one of two ways. Core accretion of gas onto a
$\sim$10 M$_\oplus$ solid core operates within $\sim$1-10 AU
\citep[]{Pollack:96,Lissauer:07}, while rapid collapse by gravitational instability
may occur beyond a few tens of AU \citep[]{Boss:00,Gammie:01,Boley:09}. Neither
process is thought to form in situ the gaseous hot Jupiters seen on short period orbits
\citep[e.g.][]{Mayor:95}, so some form of migration from their formation locations is
required to explain their existence.

These migration pathways are poorly constrained. Migration may occur via the planet's
interaction with, and transport through, the protoplanetary disk
\citep[e.g.][]{Papaloizou:07}, but could occur later via scattering or secular
interactions of the planet with other massive planetary or stellar components in the
system \citep[e.g.][]{Rasio:96,Fabrycky:07}. A significant number of large spin-orbit
misalignments observed in hot Jupiter systems initially supported the role of migration
by scattering phenomena \citep[e.g.][]{Winn:2010}, but such misalignments may
also be caused by planet migration through disks that are themselves misaligned
\citep[]{Batygin:12,Crida:14}. Consequently, observed dynamical properties of hot Jupiters have been unable to conclusively constrain their migration pathways. 

In the present work, we suggest that chemical abundances of hot Jupiters have the potential to constrain their migration mechanisms. Following initial constraints on elemental abundance ratios in hot Jupiters \citep[e.g.][]{Madhusudhan:11a,Madhusudhan:12}, several studies in recent years have attempted to explain high C/O ratios in their atmospheres based on the local chemical conditions in the protoplanetary disk where the planets formed \citep{Madhusudhan:11b,Oberg:11,Mousis:11,Moses:13,Ali-Dib:14}. We investigate the influence of both the formation conditions of hot Jupiters as well as their subsequent migration to close-in orbits on their elemental abundances. 

\section{Modelling Accretion and Migration}
\label{sec:method}

We simulate the formation and migration of hot Jupiters under different scenarios and 
keep track of their chemical compositions. Our models comprise four 
key components: (i) a time-dependent protoplanetary disk, (ii) planet growth by 
accretion of solids and gas, both under core-accretion and gravitational instability,  
(iii) planet migration, both within and without the disk, and (iv) chemical evolution of the planet.

\subsection{Protoplanetary Disk}
\label{sec:disk}

The protoplanetary disk comprises solid and gaseous components. For the gaseous
component, we use the time-dependent viscous thin disk model of \citet{Chambers:09}. We
assume an initial gas-to-dust ratio of 100 to derive the surface density of solids in the
disk, increasing this solid surface density by a factor of two at the ``snow line''
($T_{\rm disk}=170$ K) to account for the additional solids from water ice. We fix the
stellar mass to $1M_\odot$, the viscosity $\alpha$ to 0.01, the initial disk mass to
$0.15M_\odot$, and the simulation start time to 1Myr to allow for prior embryo growth. The
disk accretion rate is the steady-state solution $\dot{M}_{\rm disk}=3 \pi \nu
\Sigma_{\rm g}$.

\begin{figure}[t]
\centering
\includegraphics[width = 0.5\textwidth]{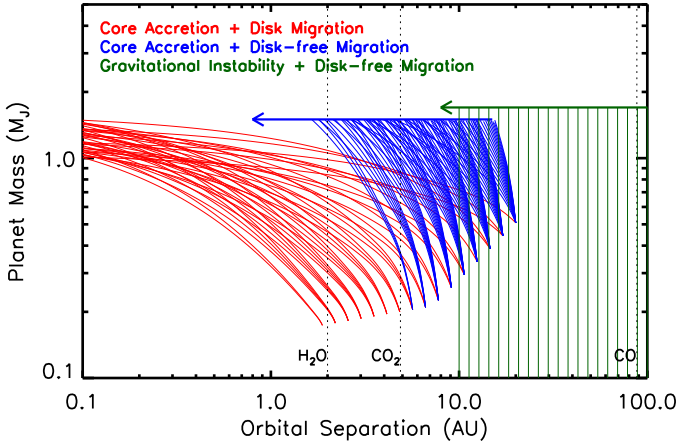}
\caption{Evolutionary tracks showing the formation and migration of hot Jupiters under
  three different conditions. The planets start at the lower-right and move to the upper-left. 
  Only a representative population is shown in each case. The
  red curves show evolution of hot Jupiters that formed via core accretion and migrated
  through the disk. The blue curves show evolutionary tracks of planets that formed by
  core accretion but migrated through disk-free mechanisms, e.g. dynamical scattering. The green curves show tracks of hot Jupiters that formed via gravitational instability, and migrated disk-free. The vertical dotted lines show snow lines of H$_2$O, CO$_2$, and CO for a
  representative disk.} 
\label{fig:tracks}
\end{figure}

\subsection{Growth of Giant Planets}

We consider giant planet formation by both core accretion and gravitational
instability. Our core accretion simulations of hot Jupiters follow a population 
synthesis approach \citep[e.g.][]{Alibert:05,Benz:14}. Our simulations begin with the opening of a
gap in the disk, i.e. at the onset of Type II migration \citep[e.g.][]{Papaloizou:07},
which sets the initial planet mass. Gas accretion on to the planet in the Type II regime
after gap opening is set using the scaling relation from \cite{Veras:04}. We further
multiply this rate by an efficiency factor $f_{\rm accr}$ to allow for the possibility
that material flows past the planet at a greater rate than implied by this
prescription \citep[see also][]{Lubow:06}.

Growing planets also accrete solid material from the disk, for which we use the solid
accretion rate given by \cite{Alibert:05}, including the capture efficiency factor that
allows for planetesimals to be accreted inefficiently when the planet's escape velocity
is large \citep{Ida:04}. The solid material accreted onto the planet is removed from an
annular disk region 8 Hill radii wide centered on the planet of mass $M_{\rm pl}$ at
semi-major axis $a_{\rm pl}$.

We model gravitational instability following the approach of
\citet{Helled:08,Helled:09}. The initial masses of such planets is in the gas giant
regime \citep{Boley:10}, so significant subsequent growth (or disruption) cannot occur
for these planets to be detected as $\sim$1$M_{\rm Jup}$ hot Jupiters (and not brown
dwarfs). The migration is therefore either very rapid relative to further growth
(i.e. effectively $f_{\rm accr} \approx 0$), or occurs after dispersal of the gas
disk. Most of the planet mass is therefore acquired in-situ, with some solids
subsequently accreted after gravitational collapse to reach the final planet
mass. \citet{Helled:08} neglect gravitational focussing, which probably underestimates
the solid mass accreted so we compute models with this prescription and alternatively
assume that all solids within 8 Hill radii of the planet are accreted. These extremes
likely cover the actual mass in solids accreted in such a scenario. Once the mass of
solids acquired by the planet is determined, the mass in gas is just the difference
between the total mass of the planet (fixed to near 1$M_{\rm Jup}$) and the mass in
solids.

\begin{deluxetable*}{l l l l}
\tablewidth{0pt} 
\tabletypesize{\scriptsize}
\tablecaption{Volume mixing ratios of chemical species in the disk} 
\tablehead{\colhead{Species} & \colhead{T$_{\rm cond}$ (K) \tablenotemark{a}} & \colhead{Case 1\tablenotemark{b}: X/H} & \colhead{Case 2\tablenotemark{c}: X/H}}
\startdata  
CO  & 20  & 0.45$\times$ C/H (0.9$\times$ C/H for T $<$ 70 K)& 0.65 $\times$ C/H \\ 
CH$_4$  & 30  & 0.45$\times$ C/H (0 for T $<$ 70 K) & 0 \\ 
CO$_2$  & 70  & 0.1$\times$ C/H (0 for T$<$70 K) & 0.15 $\times$ C/H \\  
H$_2$O  & 170  & O/H - (3$\times$ Si/H + CO/H + 2$\times$ CO$_2$/H)& O/H - (3$\times$ Si/H + CO/H + 2$\times$ CO$_2$/H) \\ 
Carbon grains & 150 & 0 & 0.2 $\times$ C/H \\
Silicates & 1500 & Si/H & Si/H
\enddata
\tablenotetext{a}{The condensation temperatures for CO, CH$_4$, CO$_2$, and H$_2$O are adopted from \citet{Mousis:11}, and those for silicates and carbon grains are adopted from \citet{Oberg:11}.}
\tablenotetext{b}{Volume mixing ratios of the various species adopted based on theoretical computations for a solar composition disk \citep{Woitke:09}. For temperatures higher than the CO$_2$ condensation curve, i.e. at smaller orbital separations, the densities can be high enough for methane (CH$_4$) to be abundant in chemical equilibrium. Non-equilibrium chemistry could also lead to some CO production. CO$_2$ is less abundant than CO or CH$_4$, depending on the temperature regime \citep{Madhusudhan:11c}. We have assumed a CO$_2$ fraction to be $10\%$ of C, and partitioned the remaining C between CO and CH$_4$. Farther out in the disk, beyond the CO$_2$ snow line, CO contains the majority of C and CH4 is negligible \citep{Woitke:09}.}
\tablenotetext{c}{Volume mixing ratios of the various species following \citep{Oberg:11}who  adopted values based on ice and gas observations in protoplanetary environments). This set of compositions contains carbon grains leading to more solid carbon.} 
\label{tab:chem}
\end{deluxetable*}

\subsection{Giant Planet Migration}

We consider giant planet migration both through interactions with the gas disk, and via
dynamical interactions with a third body. We model inward Type
II migration of a planet through the disk, which occurs for planets with more than a few
tens of Earth masses \citep[e.g.][]{Papaloizou:07}. The planet accretes gas as it
migrates, so the observable planet atmosphere is dominated by the mass accreted once Type
II migration begins and earlier phases are of minor concern here (and we do not model
them). The relative rates of gas/solid accretion and migration set the locations in the
disk where the planet's envelope came from, and hence the observable chemical properties
of the planet.

We adopt the Type II migration rate used by \citet{Ida:04}, though other migration
prescriptions are possible \citep[e.g.][]{Mordasini:09}. Our model accounts for the
possibility of different migration prescriptions by varying the accretion efficiency via
the factor $f_{\rm accr}$. Given the prescriptions for gas accretion and migration of a
giant planet, an expression for planet growth during migration, which is independent of
the disk properties, is
\begin{equation}\label{eq:dmda}
  \frac{d \log M_{\rm pl}}{d \log a_{\rm pl}} \sim - 5f_{\rm accr}\sqrt{a_{\rm pl}/30{\rm AU}} \, .
\end{equation}
At large radii migration is slow relative to growth, but increases at small radii
(i.e. $1$ AU). This growth can be visualized in a plot of $M_{\rm pl}$ vs. $a_{\rm pl}$
as shown in Fig. \ref{fig:tracks}, with curves following the formation of planets with
different initial conditions and $f_{\rm accr}$ \citep[see also][]{Mordasini:09,Rice:13}.

Migration via scattering or secular perturbations (i.e. 'disk-free' migration) is treated by truncating planet growth at the target mass of 1~$M_{\rm Jup}$. For core accretion and disk-free migration, the planet evolves via Type-II migration until it reaches 1~$M_{\rm Jup}$ following which we assume the growth is truncated due to a scattering event or disk dispersal. For gravitational instability we simply assume in situ formation prior to scattering. In either case the planetary chemical composition is fixed when 1~$M_{\rm Jup}$ is reached, and the semi-major axis decreased to 0.1 AU without further accretion. In reality, the planet may continue to accrete during disk-free migration, the chemical implications of which in our models are similar to those involving Type-II migration followed by scattering, as discussed in section~\ref{sec:discussion}. 

\begin{figure}[b]
\centering
\includegraphics[width = 0.5\textwidth]{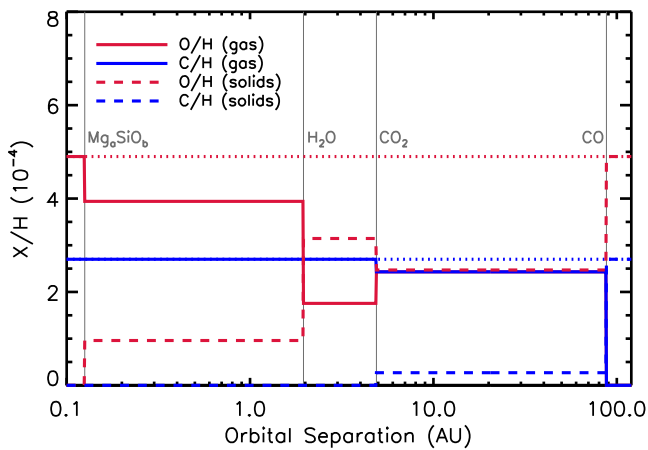}
\caption{Partitioning of oxygen and carbon in the mid-plane of a model protoplanetary
  disk. The molecular mixing ratios correspond to Case 1 of Table~\ref{tab:chem} for illustration. The solid red (blue) curves show the volume mixing ratio of oxygen (carbon) in the gas, and the dashed curves show the corresponding volume mixing ratios in solid planetesimals, in the form of ices or
  rocks. The dotted lines show the total abundances of O and C in gas and solids, which
  equal solar abundances. The condensation fronts of different molecular species are
  shown in the thin vertical lines.} 
\label{fig:chem}
\end{figure}

\subsection{Chemistry}

The dominant O and C bearing species in the disk mid-plane are H$_2$O, CO$_2$, CO, CH$_4$,
silicates, and graphite grains. The volume mixing ratios of these species for a solar
composition disk are derived based on chemical equilibrium in the disk mid-plane
\citep{Woitke:09} as well as from observations of protoplanetary disks
\citep{Oberg:11,Draine:03, Pontoppidan:06}; our adopted values are shown in Table~\ref{tab:chem}. The
snow lines of H$_2$O, CO$_2$, and CO govern the apportionment of C and O in gaseous versus
solid phases, apart from minor fractions of C and O in refractory condensates such as
silicates and graphite grains \citep{Pontoppidan:06}. The mid-plane elemental composition of a representative solar abundance disk is shown in Fig.~\ref{fig:chem}. For separations closer to the star
than the H$_2$O snow line, the O and C are predominantly in gas phase with minor
quantities in silicates and carbides. For larger separations, O is progressively depleted
from the gas and moved into solid ices following the various snow lines starting with
H$_2$O. C is predominantly in gas phase until the CO$_2$ and CO snow lines
\citep[also see e.g.,][]{Oberg:11}.

We perform the calculations for two assumptions of chemistry, one motivated by
theoretical calculations of chemistry in H$_2$-rich environments \citep{Woitke:09,Madhusudhan:11c}, and another based on observations of ice and gas in protoplanetary environments \citep{Oberg:11}.  In
Table~\ref{tab:chem}, we express the volume fractions of species in terms of elemental volume
fractions for solar composition \citep{Asplund:09}: O/H $= 4.9 \times 10^{-4}$, C/H $=
2.7 \times 10^{-4}$, and Si/H $= 3.2\times 10^{-5}$. At each time-step of the planet's growth and migration, the mass fractions of the various chemical species in solids and gas accreted by the planet are determined based on the planet's feeding zone, mid-plane temperature, and net mass accreted in solids and gas. 

\subsection{Accretion/migration Scenarios}
\label{sec:scenarios}
We now combine the above model components to investigate various formation-migration pathways 
that can lead to hot Jupiters. Our goal is to form planets with $M_{\rm pl}\approx 1 M_{\rm Jup}$ and 
$a_{\rm pl} \approx 0.1$ AU around sun-like stars with solar abundances. We consider three broad 
formation-migration scenarios as shown in Fig. \ref{fig:tracks}, and explore a wide range of initial conditions in each: (a) planets formed via core accretion with initial separations at $\sim 1-20$ AU and migrated to 0.1 AU through the disk, (b) planets  formed via core-accretion but migrated disk-free, (c) planets formed via gravitational instability in situ but migrated disk-free. We do not model gravitational instability followed by disk migration since this scenario results in planets more massive than the 1~$M_{\rm Jup}$ planets considered here. Other modeling aspects for future considerations are discussed in section~\ref{sec:discussion}. 

\section{Planetary Chemical Compositions}
\label{sec:cxo}
\begin{figure}[t]
\centering
\includegraphics[width = 0.5\textwidth]{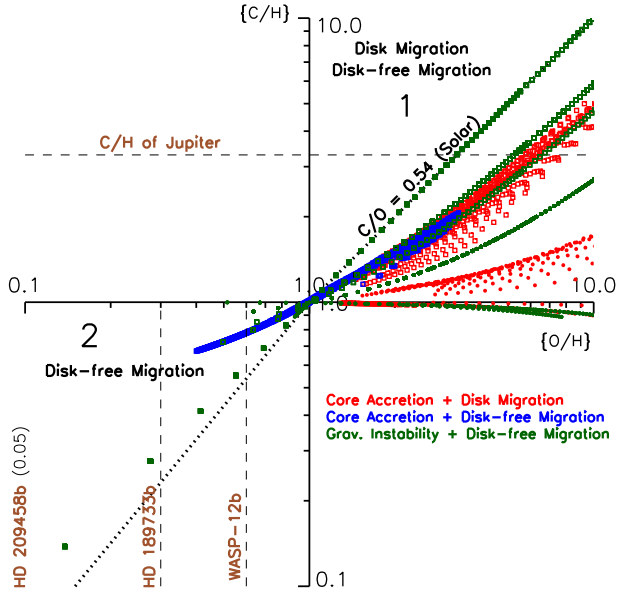}
\caption{Oxygen and Carbon abundances of hot Jupiters formed through different
  formation-migration pathways. The x and y axes denote O/H and C/H volume mixing ratios
  in the planetary envelope normalized to solar values. The red, blue, and green symbols
  show hot Jupiters that formed via different combinations of formation and migration
  mechanisms as shown in the legend. The circles and squares indicate different
  prescriptions for chemistry (cases 1 and 2, respectively, in Table~1). The vertical dotted lines in
  the lower left quadrant show O/H abundances derived from the sub-solar H$_2$O abundances 
  reported for three hot Jupiters by \citet{Madhusudhan:14}; the highest value within the uncertainties is adopted here. See section~\ref{sec:discussion}.}
\label{fig:COplane}
\end{figure}

For each of the three scenarios discussed in \ref{sec:scenarios} and 
illustrated in Fig. \ref{fig:tracks}, the resulting elemental abundances in terms of the O/H and C/H 
ratios are shown in Fig. \ref{fig:COplane}. 

For Jovian-mass hot Jupiters formed via core accretion and disk migration, the oxygen 
abundances range from solar to over $10\times$ solar values, whereas the carbon
abundances range between solar and $\sim$5$\times$ solar values. The
solids in this separation range ($\sim$1-20 AU) are dominated by O in the 
form of H$_2$O-ice between the H$_2$O and CO$_2$ snow lines, with
additional contribution due to CO$_2$-ice beyond the CO$_2$ snow line, while some O is 
also present in silicates at all separations. The amount of carbon in planetesimals  
is sensitive to its assumed abundance in grains, as shown in Table~\ref{tab:chem}. 
Assuming C is present only in volatiles (Case 1 in Table~\ref{tab:chem}), C in solids is  
only due to CO$_2$-ice and leads to a C/H of $\sim$1-1.5$ \times$ solar. On the
other hand, following Case~2 if C is present as graphite grains \citep{Oberg:11,Draine:03}, 
C abundances up to $\sim$5$\times$ solar are possible. If carbon-based volatiles, e.g. CH$_4$ and/or CO, are locked into H$_2$O-ice forming clathrates \citep{Mousis:11} the C abundance is
further increased. Irrespective of the form of C, the C/O ratio is always
sub-solar. Assuming Jupiter's formation by core accretion and using its observed 
C/H \citep{Atreya:10}, as shown in Fig. \ref{fig:COplane}, we predict its O/H to be 
$\sim$5-8$\times$ solar which is consistent with previous estimates \citep{Mousis:12}. Thus, hot 
Jupiters formed via core accretion and disk migration could contain solar or super-solar C/H and O/H, 
but would have sub-solar C/O, and would occupy the lower half of the top-right quadrant 
(region 1) of the C-O plane in Fig. \ref{fig:COplane}. 

Hot Jupiters that formed via core accretion but underwent disk-free migration,
can have super-solar or sub-solar abundances, as shown in Fig. \ref{fig:COplane}. Planets 
that formed closer-in accrete solids more efficiently, resulting in super-solar abundances, 
before accretion is halted due to scattering. These planets also occupy region 1 of the 
C-O plane in Fig. \ref{fig:COplane}. Planets forming farther 
out tend to have sub-solar elemental abundances, due to an enhanced ability to eject rather than 
accrete at large separations as discussed in section~\ref{sec:method}, and super-solar C/O ratios. 
For planets originating between the H$_2$O and CO$_2$ snow lines, the O/H is sub-solar whereas 
the C/H is nearly solar. Those originating beyond the CO$_2$ snow line result in sub-solar C/H as well 
as sub-solar O/H, thereby occupying the bottom-left quadrant of the C-O planet (region 2). The presence 
of C in graphite can further lower the C/H ratio for such planets, while still retaining a super-solar C/O ratio.

Planets that originated at large orbital separations ($\gtrsim$20 AU), are likely to have formed in situ via gravitational instability followed by disk-free migration. As shown in Fig. \ref{fig:COplane}, such planets can have super-solar as well as sub-solar O and C while their C/O ratios depend on their formation location. For planets forming beyond the CO snow line ($\sim$80 AU), the C/O ratio is close to solar independent of the overall metallicity as the important volatiles are all in solid form. For planets forming in this region the metallicity and C/O ratio are inversely correlated; for super-solar abundances the C/O
ratio is sub-solar whereas for sub-solar abundances the C/O ratio is super-solar, both
governed by the location of the planet with respect to the H$_2$O and CO$_2$ ice-lines. 

Overall, our results suggest a preliminary framework for placing chemical constraints on migration 
mechanisms of hot Jupiters. Super-solar O and C abundances are unlikely to yield 
conclusive constraints as they can result from a variety of formation-migration pathways
However, sub-solar O and C abundances and super-solar C/O ratios could be indicative of 
disk-free migration of hot Jupiters irrespective of whether they formed by core accretion 
or gravitational instability.  

\section{Discussion}
\label{sec:discussion}
Recent observations suggest sub-solar H$_2$O abundances in several hot Jupiters, e.g. HD~189733b, HD~209458b, and WASP-12b \citep{Madhusudhan:14,Stevenson:14}. It is unclear if clouds are responsible for the muted H$_2$O features \citep{Madhusudhan:14,McCullough:14}, and the low H$_2$O may indeed represent the bulk atmospheric composition. A bulk sub-solar H$_2$O abundance could result either from a sub-solar bulk metallicity and/or a C/O$\geq$1. If the bulk metallicity is sub-solar, the limits on the O/H in these planets are shown in Fig~\ref{fig:COplane},  which would suggest disk-free migration as a potential explanation for their origins. Alternatively, the C/O could be $\geq$1 as has been suggested for WASP-12b \citep{Madhusudhan:11a,Stevenson:14}, but the C/H ratio is presently unconstrained but observable in the future. 

Future studies will need to investigate in detail several aspects of the present population study \citep[also see][]{Benz:14}. In principle, a rigorous simulation of the chemical evolution of a planet while forming and migrating would include hydrodynamic disk evolution along with N-body planetesimal dynamics and self-consistent chemistry. Such next-generation models may predict a wider dispersion in elemental abundances than shown in Fig.~\ref{fig:COplane}. 
We have assumed that no accretion happens after the onset of disk-free migration via a scattering event, while in reality the disk may still be present and accretion may continue especially at short orbital separations. In such cases, a higher elemental enhancement is possible, especially in oxygen, and some of the systems from region 2 of Fig.~\ref{fig:COplane} may move into region 1 analogous to our models with Type-II migration followed by scattering. Conversely, the influence of `planetesimal shepherding' by the inward migrating hot Jupiter \citep{Fogg:07a,Fogg:07b} may result in lower solid accretion than we have assumed. Overall, however, hot Jupiters with sub-solar C/H and O/H will be hard to explain via disk migration as they would need to (a) form at large orbital separations beyond the CO$_2$ snow line ($\gtrsim$5 AU), and (b) migrate through the disk all the way to close-in orbits without accreting any significant mass in solid planetesimals.  

We have assumed that the bulk elemental abundances in the protoplanetary disk mimic those of the host star, in this case solar. However, future observations of protoplanetary disks could help provide a range of plausible chemical abundances and disk properties that could be used as inputs in modeling studies. New studies would also need to investigate the influence of the accreted solids on the atmospheric compositions of hot Jupiters. For hot Jupiters, the temperatures in the deep atmosphere, at $\sim$100-1000 bar, can be $\gtrsim$2000 K \citep{Burrows:08} which is hotter than the sublimation temperatures of all ices and silicates. The interiors are even hotter and convective, because of which we have assumed the elements in solids accreted are well mixed in the planetary envelope. 

Overall, our study presents a first attempt at investigating chemical constraints on the
migration mechanisms of hot Jupiters. Future measurements of precise elemental abundances
in hot Jupiter atmospheres and concomitant advances in theoretical work extending the
ideas presented here might lead to new insights on the migration of hot Jupiters, which
have thus far remained enigmatic based on dynamical constraints alone.

\acknowledgements{GMK acknowledges support from the European Union through ERC grant number 279973.}

\vspace{5pt}


\end{document}